\documentclass{article}

\usepackage{graphicx}
\usepackage{epsfig}
\usepackage{amsfonts}

 \usepackage{amssymb}
\usepackage{amsmath}
\usepackage{hyperref}
\usepackage{graphicx}       % Include figure files
\usepackage{dcolumn}        % Align table columns on decimal point
\usepackage{bm}             % bold math
\usepackage{amssymb}
\usepackage{amstext}

\date{\today}

\newcommand{\insertplot}[5]{\begin{figure}
 \hfill\hbox to 0.05in{\vbox to #5in{\vfill
 \inputplot{#1}{#4}{#5}}\hfill}
 \hfill\vspace{-.1in}
 \caption{#2}\label{#3}
 \end{figure}}
 \newcommand{\inputplot}[3]{
 \special{ps: plotfile #1}
}
\newcounter{fig}

\newcommand{\ee}{\end{equation}}
\newcommand{\eea}{\end{eqnarray}}

\newcommand{\eps}{\epsilon}

\newcommand{\beq}{\begin{equation}}
\newcommand{\eeq}{\end{equation}}

\begin{document}

\title{
\Large{\bf 
Probing the 
Ellis-Bronnikov wormhole  geometry 
\\
with  a scalar field: clouds, waves and Q-balls
}
}
 \vspace{1.5truecm}

\author{
{\large }%$^{\ddagger}$
{\ Jose Luis Bl\'azquez-Salcedo}$^{1}$,
{\ Marina-Aura Dariescu}$^{2}$, 
{\ Ciprian Dariescu}$^{2}$,
\\
{ Eugen Radu}$^{3}$
and
{ Cristian Stelea}$^{4}$
\\
\\
$^{1}${\small  Departamento de F\'isica Te\'orica and IPARCOS, Universidad Complutense de Madrid, E-28040 Madrid, Spain}
\\
$^{2}${\small  Faculty of Physics, ``Alexandru Ioan Cuza" University of Iasi
Bd. Carol I, no. 11, 700506 Iasi, Romania}
\\ 
$^{3}${\small
Departamento de Matem\'atica da Universidade de Aveiro and}
\\
{\small  Centre for Research and Development  in Mathematics and Applications (CIDMA),} 
\\ 
{\small    Campus de Santiago, 3810-183 Aveiro, Portugal}
\\
$^{4}${\small Department of Exact and Natural Sciences, Institute of Interdisciplinary Research,}
\\
{\small  ``Alexandru Ioan Cuza University" of Iasi,  Bd. Carol I, no. 11, 700506 Iasi, Romania} 
}

\maketitle

%\date{\today}
%\pacs{ }
\begin{abstract}
 The Ellis-Bronnikov solution  
provides a simple toy model for the study of various aspects
of wormhole physics.
In this work we solve the Klein-Gordon equation in this background 
and find an exact solution in terms of Heun's function.
This may describe 'scalar clouds'
($i.e.$ localized, particle-like configuration) or scalar waves.
However, in the former case, the radial derivative of the scalar field
is discontinuous at the wormhole's throat (except for the spherical case).
This pathology is absent for a suitable  scalar field self-interaction,
and we provide evidence for the existence of spherically symmetric
and spinning Q-balls in a  Ellis-Bronnikov wormhole background.
\end{abstract}

%\tableofcontents

%%%%%%%%%%%%%%%%%%%%%%%%%%%%%%%%%%%%%%%%%%%%%%%%%%%%%%%%%%%%%%%%%%%%%%%%%%%%%%%%%%%
\section{Introduction}
%%%%%%%%%%%%%%%%%%%%%%%%%%%%%%%%%%%%%%%%%%%%%%%%%%%%%%%%%%%%%%%%%%%%%%%%%%%%%%%%%%%

The study of (classical) solutions of a field theory in a given spacetime background
is an important step towards various more complicated studies.
For example, in the field quantization
  one starts usually with the field modes construction \cite{Birrell:1982ix};
also, in black hole physics, a number of  
'no-hair' theorems can be established  at the level of matter field equations, 
without the use of gravity equations 
\cite{Bekenstein:1996pn},
\cite{Herdeiro:2015waa}. 

A particularly interesting type of spacetime geometry which is allowed by the Einstein's field equations is
provided by the (Lorentzian) traversable wormholes (WHs). 
These are 'topological handles'  connecting separated regions of a single
Universe or “bridges” joining two different spacetimes.
As such, they are (at least) interesting solutions of the General Relativity
(with its various extensions), being a useful tool to probe the limits of the theory.
The study of WHs started in with the work of Flamm  in 1916 
\cite{Flamm} together with the 
  Einstein and Rosen paper in 1935
	\cite{Einstein-Rosen}. 
After several decades of (relatively) slow progress (see, however, Wheeler's work 
\cite{Wheeler:1957mu},
\cite{Wheeler}),
 a fresh interest in the topic of  WHs has been reawaken by the work of
Morris and Thorne \cite{Morris-Thorne}, the field branching off into diverse directions.
Among other results, the work \cite{Morris-Thorne} has clarified that  
some form of exotic matter violating the energy conditions 
is necessary in order to keep the throat of the WH open.
 
One of the simplest (and an early) example of traversable WHs in General Relativity
has been found in 1973 by Ellis
\cite{Ellis} and Bronnikov \cite{Bronnikov}.
 This is a solution of the Einstein equations minimally coupled with a massless phantom scalar field 
 ($i.e.$ with a wrong sign in front of its kinetic term).
The Ellis-Bronnikov solution is an archetypal example of a  WH geometry, with a large number of papers
investigating its properties (here we mention only the results in \cite{Gonzalez:2008wd}
establishing 
that this configuration is unstable).

In the context of this study, the Ellis-Bronnikov geometry is of interest 
mainly
because of its simple form,
which allows for a more systematic study of the solutions of a field theory model. 
For simplicity, in this work we shall consider the simplest case of a massive scalar field, 
which may possess a self-interacting
potential.

The initial motivation for this study came from this simple observation
that removing a sphere from  Minkowski spacetime 
allows for everywhere regular scalar 
multipoles, with a finite total mass.
For concreteness, 
let us consider the Laplace equation for a static scalar field, $\nabla^2 \Phi=0$.
In flat spacetime, its  general solution  is
 described by a multipolar expansion,
with $\Phi=\sum_{\ell,m} R_\ell(r) Y_\ell^m(\theta,\varphi)$,
where $Y_\ell^m$ are spherical harmonics and
 $R_\ell(r)=c_1 r^\ell+c_2/r^{\ell+1}$
 ($(r,\theta,\varphi)$ being the usual spherical coordinates). 
Thus, any non-trivial solution diverges either at the origin or at infinity. 
However, the situation changes if we restrict the domain of existence of the field outside a sphere of radius
 $r_B$, and take $c_1=0$.
Then any field mode is finite, with a nonzero total mass.
 
\medskip 

In some sense, a WH provides an explicit realization of this scenario; 
since the two sphere possesses
a minimal nonzero size, and one can predict the existence in this case
of  scalar clouds  
($i.e.$ particle-like, localized configurations with finite mass).
Indeed, this is confirmed by the analysis in Section 3 of this paper, where we 
find  closed form solutions of the Klein-Gordon equation in a Ellis-Bronnikov 
WH background which can be interpreted as 'scalar clouds'.
However,
they fail generically to satisfy the Klein-Gordon equation at the throat, 
with a discontinuity
in the first radial derivative of the field,
the only exception being the spherically symmetric configuration.
Physically reasonable solutions are found to exist for scalar waves, only.

When turning on the scalar field self-interaction, 
we find in Section 4  that this cures the pathological behaviour of the  scalar
clouds at the throat. 
Focusing on a complex massive scalar field with quartic
plus hexic self-interactions,
we find 
numerical evidence is given for the existence of 
spherically symmetric and axially symmetric spinning Q-ball-type solutions in 
 Ellis-Bronnikov WH background.

%%%%%%%%%%%%%%%%%%%%%%%%%%%%%%%%%%%%%%%%%%%%%%%%%%%%%%%%%%%%%%%%%%
\section{The model}
%%%%%%%%%%%%%%%%%%%%%%%%%%%%%%%%%%%%%%%%%%%%%%%%%%%%%%%%%%%%%%%%%%

We consider the action for a complex scalar field $\Phi$
with a self-interaction potential $U$
\begin{equation}
\label{action}
S=-\int \left[ 
   \frac{1}{2} g^{\mu\nu}\left( \Phi_{, \, \mu}^* \Phi_{, \, \nu} + \Phi _
{, \, \nu}^* \Phi _{, \, \mu} \right) + U( \left| \Phi \right|) 
 \right] \sqrt{-g} d^4x
\ , 
\end{equation}
where  
the asterisk denotes complex conjugation
and
$\Phi_{,\, {\mu}} ={\partial \Phi}/{ \partial x^{\mu}}$.

Variation of (\ref{action}) with respect to the scalar field
leads to the (non-linear) Klein-Gordon  equation
\begin{equation}
\label{KG}
\nabla^2\Phi= \frac{\partial U}{\partial\left|\Phi\right|^2}\Phi \ .
\end{equation}
%where $\Box$ represents the covariant d'Alembert operator.
%
%
The stress-energy tensor $T_{\mu\nu}$ of the scalar field is
\begin{eqnarray}
\label{tmunu} 
T_{\mu \nu} 
=
\left( 
\Phi_{, \, \mu}^*\Phi_{, \, \nu}
+\Phi_{, \, \nu}^*\Phi_{, \, \mu} 
\right ) 
-g_{\mu\nu} \left[ \frac{1}{2} g^{\alpha\beta} 
\left( \Phi_{, \, \alpha}^*\Phi_{, \, \beta}+
\Phi_{, \, \beta}^*\Phi_{, \, \alpha} \right)+U(\left| \Phi \right|)\right]~.
\end{eqnarray}

In the above relations  $g_{\mu \nu}$
is taken to be the metric tensor of the Ellis-Bronnikov solution,
with the following parametrization
\begin{eqnarray}
\label{metric}
ds^2= dr^2 + (r^2 + r_0^2 ) \left( d \theta^2 + \sin^2 \theta d \varphi^2 \right) - dt^2~,
\end{eqnarray} 
 $\theta$ and $\varphi$
being spherical coordinates with the usual range, 
while $r$ and $t$ are the radial and time coordinates, respectively.
This geometry consists in two different regions 
$\Sigma_\pm$;
the `up' region ($\Sigma_+$) is found for $0<r<\infty$,
while the 'down' region ($\Sigma_-$) has $-\infty<r<0$.
These regions are joined at $r=0$, which is the position of the 
spherical throat,
which is a minimal
surface of area $4\pi r_0^2$.

For the geometry (\ref{metric}),
the equation (\ref{KG})
takes the form
\begin{eqnarray}
\label{KG1}
\frac{\partial^2 \Phi}{\partial r^2} + \frac{2r}{r^2 + r_0^2} \frac{\partial \Phi}{\partial r} + \frac{1}{r^2+r_0^2} \left[
\frac{\partial^2 \Phi}{\partial \theta^2} + \cot \theta \frac{\partial \Phi}{\partial \theta} +\frac{1}{\sin^2 \theta} \frac{\partial^2 \Phi}{\partial \varphi^2}  \right]
- \frac{\partial^2 \Phi}{\partial t^2} 
-\frac{\partial U}{\partial\left|\Phi\right|^2 }\Phi  = 0.
%- \mu^2 \Phi = 0.
\end{eqnarray} 

The model is invariant under the global phase transformation
$\Phi \to \Phi e^{i\alpha}$,  
 leading to the conserved current
 \begin{eqnarray}
\label{scalar-current}
j^{\mu}=-i\left[(\Phi)^* \partial^\mu\Phi+\Phi \partial^\mu\Phi^*\right]\ , \qquad \nabla_\mu j^{\mu}=0 \ .
 \end{eqnarray}
This implies the existence of a 
 conserved  Noether charge (corresponding to particle number),
 which is the integral of $j^t$ on spacelike slices,
\begin{eqnarray}
\label{Q}
Q_{\pm} =  \int_{\Sigma_\pm } d^3 x \sqrt{-g}~ j^t.
 \end{eqnarray}
Morover, for  particle-like solutions (scalar clouds)
one can assign a mass  
in both 'up'and 'down' regions  
\begin{eqnarray}
\label{mass}
E_{\pm} = -\int_{\Sigma_\pm } d^3 x \sqrt{-g}~ T_t^t .
 \end{eqnarray}
One should note that these masses $E_{\pm}$ are assigned to the scalar field itself and they do not refer to the gravitational masses that can be computed for the Ellis-Bronnikov WH background in each asymptotic region.

%%%%%%%%%%%%%%%%%%%%%%%%%%%%%%%%%%%%%%%%%%%%%%%%%%%%%%%%%%%%%%%%%%
\section{The non-selfinteracting case}
%%%%%%%%%%%%%%%%%%%%%%%%%%%%%%%%%%%%%%%%%%%%%%%%%%%%%%%%%%%%%%%%%%

%%%%%%%%%%%%%%%%%%%%%%%%%%%%%%%%%%%%%%%%%%%%%%%%%%%%%%%%%%%%%%%%%%
%\subsection{A general solution}
%%%%%%%%%%%%%%%%%%%%%%%%%%%%%%%%%%%%%%%%%%%%%%%%%%%%%%%%%%%%%%%%%%

For the case of a massive scalar field with
\begin{eqnarray}
 U( \left| \Phi \right|)=\mu^2  |\Phi|^2,
\end{eqnarray}
(where $\mu \geq 0$ is the  boson mass),
one may separate the variables as
\begin{equation}
\Phi = R_\ell(r) Y_{\ell}^m (\theta,\varphi) e^{- i \omega t} ,
\end{equation}
with 
$\omega$ the field frequency
and
$Y_{\ell}^m$ Laplace's spherical harmonics
(where $\ell \geq 0$ and $-\ell \leq m\leq \ell$).
Then, after replacing in (\ref{KG1})
one finds the following equation for the 
 radial amplitude $R_\ell(r)$
\begin{eqnarray}
\label{eqR}
\frac{1}{r^2+r_0^2} \frac{d}{dr} \left[ ( r^2 + r_0^2) \frac{dR_\ell}{dr} \right] +
 \left[ \omega^2 - \mu^2 - \frac{\ell (\ell+1)}{r^2+r_0^2} \right] R_\ell = 0~.
\end{eqnarray}
This equation possess an exact solution 
 in terms of Heun confluent functions 
\cite{Heun} 
\begin{eqnarray}
\label{gen-sol}
R_\ell(r) = c_1~H_1(r)+c_2~H_2(r)~,
\end{eqnarray}
where
\begin{eqnarray}
 &&H_1(r)={\rm  HeunC} \left[ 0 , - \frac{1}{2} , 0 , \frac{(\mu^2 - \omega^2)r_0^2}{4} , \frac{(\omega^2 - \mu^2)r_0^2}{4} 
- \frac{\ell^2+ \ell-1}{4} , - \frac{r^2}{r_0^2} \right] ~,
\\
\nonumber
&&
 H_2(r)= r_{~} {\rm HeunC} \left[ 0 , \frac{1}{2} , 0 , \frac{(\mu^2 - \omega^2)r_0^2}{4} , \frac{(\omega^2 - \mu^2)r_0^2}{4} 
- \frac{\ell^2+\ell-1}{4} , - \frac{r^2}{r_0^2} \right]~ ,
\end{eqnarray}
 $c_1,c_2$ being arbitrary constants. Let us remarks that $H_1$, $H_2$ are an even and odd functions of $r$, respectively,
such that in general $R_\ell$ has no definite parity.
 
Let us also remark that,
with the change of function
\[
R_\ell(r) = \frac{y(r)}{\sqrt{r^2+r_0^2}}~,
\]
the  eq. (\ref{eqR}) can be cast into a Schrodinger-like form
\[
-\frac{d^2y}{dr^2} + V(r) y = (\omega^2 -\mu^2)y,
~~~{\rm with}~~
V(r)=\frac{1}{r^2+r_0^2} \bigg(\ell(\ell+1) - \frac{r_0^2}{r^2+r_0^2} \bigg).
\]

%%%%%%%%%%%%%%%%%%%%%%%%%%%%%%%%%%%%%%%%%%%%%%%%%%%%%%%%%%%%%%%%%%
\subsection{The $\omega^2=\mu^2$ limit}
%%%%%%%%%%%%%%%%%%%%%%%%%%%%%%%%%%%%%%%%%%%%%%%%%%%%%%%%%%%%%%%%%%

The ${\rm HeunC}$ function possesses a  rather complicated expression;
thus, to study the solutions' properties 
we have used the software MAPLE,
calculating series expansions  and  
evaluating them numerically for various values of the parameters.

However, the 
solution (\ref{gen-sol}) greatly simplifies for $\omega^2=\mu^2$,
a case which  captures also the basic properties of the solutions
with $\omega^2 <\mu^2$.
The radial amplitude in this case reads 
\begin{eqnarray}
R_\ell (r) = c_1 {\rm L}_P( \ell ,ir/r_0) + c_2 {\rm L}_Q( \ell,ir/r_0)~,
\end{eqnarray}
 where ${\rm L}_P$ and ${\rm L}_Q$ are Legendre functions of first and second kind respectively,
the explicit form for the first values of $\ell$
being
\begin{eqnarray}
&&
\nonumber
R_0 (r) =c_1+c_2 \arctan(\frac{r}{r_0}),~~
R_1 (r) =c_1 \frac{r}{r_0} +c_2 (\frac{r}{r_0}\arctan(\frac{r}{r_0})+1),
\\
\label{ell}
&&
R_2(r)=c_1 (\frac{1}{3}+\frac{r^2}{r_0^2})
+c_2
\left(
(\frac{1}{3}+\frac{r^2}{r_0^2})\arctan(\frac{r}{r_0})+\frac{r}{r_0}
\right),~~
\\
&&
\nonumber
R_3(r)=c_1(\frac{r}{r_0}+\frac{5 r^3}{3r_0^3})
+c_2
\left(
(\frac{r}{r_0}+\frac{5 r^3}{3r_0^3})\arctan(\frac{r}{r_0})+\frac{5 r^2}{3r_0^2}+\frac{4}{9}
\right)~.
\end{eqnarray}

We are interested in localized, particle-like solutions,
with a radial amplitude which is $continuous$ and $finite$ everywhere (in particular as $r\to \pm \infty)$.
These  requirements are satisfied by the $\ell=0$ mode,
 the radial profile $R_0(r)$
approaching constant values on each asymptotic region of the wormhole.
However, one can see that for $\ell \geq 1$
 the solution necessarily diverges 
as $r\to \infty$ or as $r \to -\infty$.
Let us exemplify this for $\ell=1$, in which case
\begin{eqnarray}
\label{ell=1}
R_1\to (2c_1+c_2\pi)\frac{r}{2r_0} +O(1/r^2)~~{\rm as}~r\to \infty,~~{\rm and}~~
R_1\to (-2c_1+c_2\pi)\frac{r}{2r_0} +O(1/r^2) ~~{\rm as}~r\to -\infty.
\end{eqnarray}
The only way to cure this  pathology is to consider $R_1$ to
be the union of 
two separate solutions,
$R_1^{(+)}(r)$ 
valid for $r\geq 0$ 
(with $R_1^{(+)}\to 0$ as $r\to \infty $),  and $R_1^{(-)}(r)$ 
 (with $r\leq 0$ and $R_1^{(-)}\to 0$ as $r\to -\infty $),
which are joined at $r=0$.
That is for 
$R_1^{(+)}(r)$ 
one takes 
 $c_1=-c_2\pi/2$, 
while for 
$R_1^{(-)}(r)$  
the choice is
 $c_1=c_2\pi/2$.
This results in\footnote{Note that the solution is defined
up to an arbitrary multiplying constant which is set to one in eqs.
 (\ref{pm1}),  
(\ref{ell-gen}).}
 \begin{eqnarray}
\label{pm1}
 R_1^{(+)}(r)=1+\frac{r}{r_0}\left(\arctan(\frac{r}{r_0})-\frac{\pi}{2} \right) ,~~
 R_1^{(-)}(r)=1+\frac{r}{r_0}\left(\arctan(\frac{r}{r_0})+\frac{\pi}{2} \right) ,~~
\end{eqnarray}
The total mass of this solution is 
$E_\pm =\pi^2 r_0/3$.

The same procedure works for any value of 
$\ell$, and one finds with the following generic expression
 \begin{eqnarray}
\label{ell-gen}
R_\ell^{(+)}(r) =                   f_\ell (r)+g_\ell (r)\left(\arctan(\frac{r}{r_0})-\frac{\pi}{2} \right),~~
R_\ell^{(-)}(r) = (-1)^{\ell+1} \left(
                                    f_\ell(r)+ g_\ell (r)\left(\arctan(\frac{r}{r_0})+\frac{\pi}{2} \right) 
                               \right)
\end{eqnarray}
where 
 %\begin{eqnarray}
%
%\label{s1}
$f_\ell (-r)=(-1)^{\ell +1}f_\ell(r),$
$g_\ell (-r)=(-1)^{\ell }g_\ell(r).$
%\end{eqnarray}
such that $R_\ell^{(+)}(r)= R_\ell^{(-)}(-r)$.
The functions 
$f_{\ell}$, 
$g_\ell$ 
are polynomials, with  
 \begin{eqnarray}
\label{s2}
&&
\ell=2k:~~
 f_\ell(r)= \sum_{j=0}^{k-1}c_j (\frac{r}{r_0})^{2j+1},~~ g_\ell(r)= \sum_{j=0}^{k} \bar c_j (\frac{r}{r_0})^{2j},
\\
&&
\ell=2k+1:~~
f_\ell(r)= \sum_{j=0}^{k}d_j (\frac{r}{r_0})^{2j},~~ g_\ell(r)= \sum_{j=0}^{k} \bar d_j (\frac{r}{r_0})^{2j+1},
\end{eqnarray}
where $k=0,1,\dots$ and $c_j,\bar c_j,d_j,\bar d_j$ real coefficients.
%%%%%%%%%%%%%%%%%%%%%%%%%%%%%%%%%%%%%%%%%%%%%%%%%%%%%%%%%%%%%%%%
\begin{figure}[ht!]
%\lbfig{rhfar}
\begin{center} 
\includegraphics[height=.34\textwidth, angle =0 ]{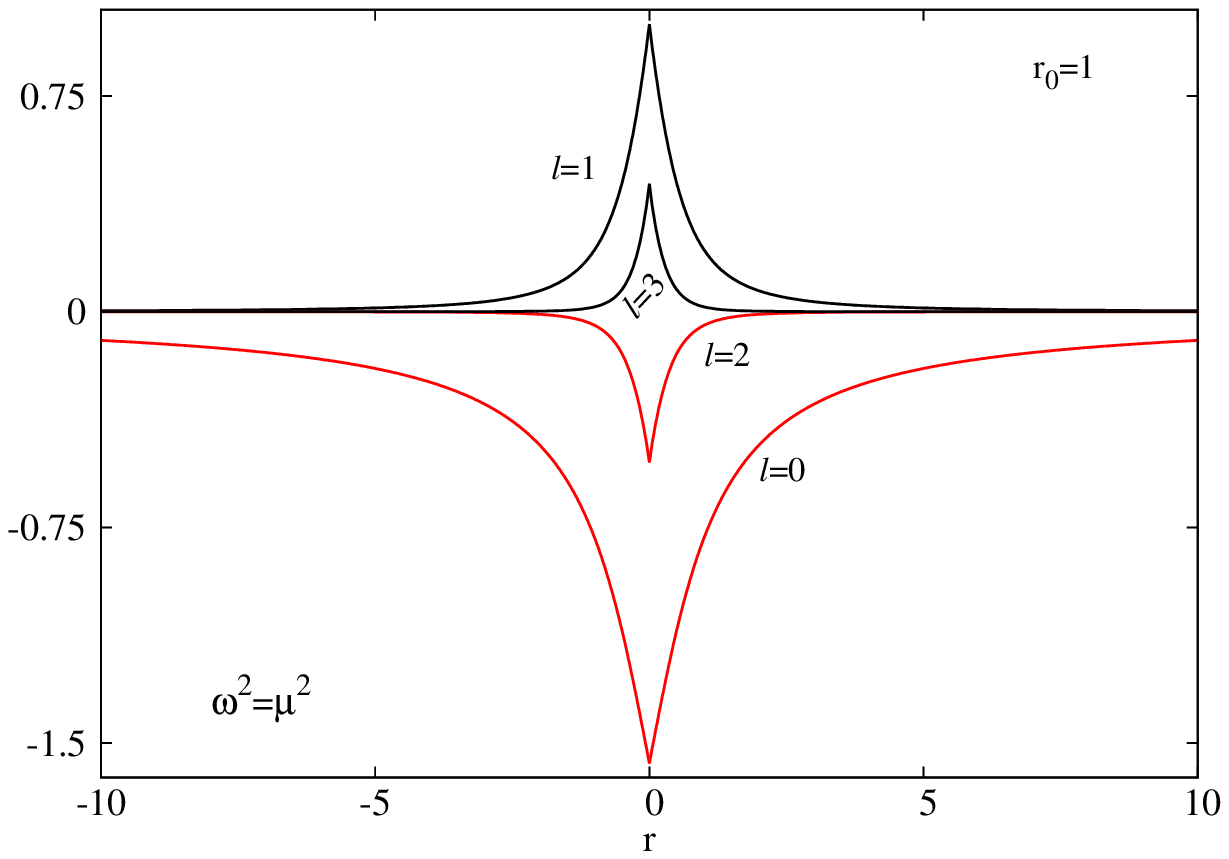}
\includegraphics[height=.34\textwidth, angle =0 ]{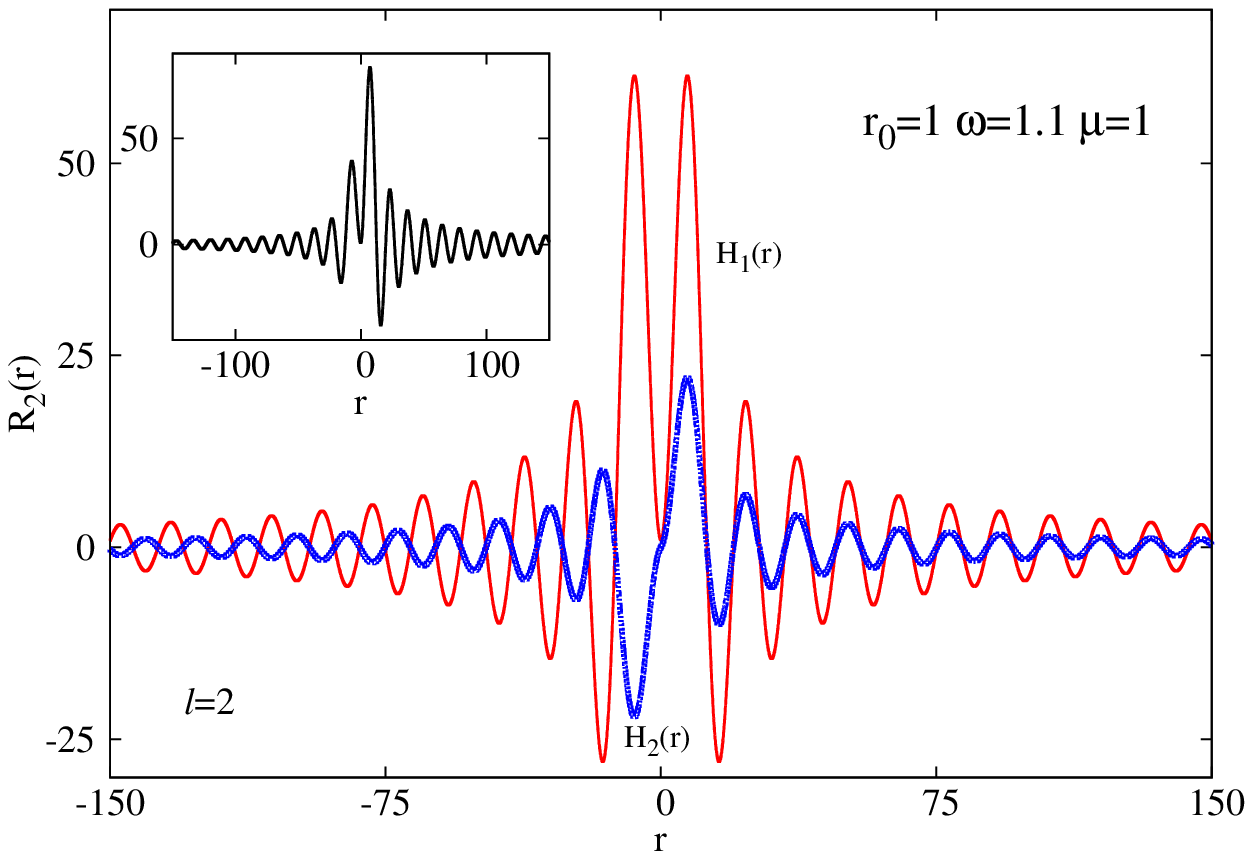}
\end{center}
\caption{
 {\it Left panel}:
The profiles of the $\ell=0,1,2,3,4$ radial amplitudes 
(\ref{ell-gen}) 
with $\omega^2=\mu^2$.
 {\it Right panel}:
A wave-like solution is shown for a $\ell=2$
configuration with $c_1=1$, $c_2=0$
($H_1(r)$)
and
$c_1=0$,
$c_2=0.9$
($H_2(r)$)
 in the near-throat expansion (\ref{nte}).
The inset shows the general solution.
}
\label{ex-sol}
\end{figure}
%%%%%%%%%%%%%%%%%%%%%%%%%%%%%%%%%%%%%%%%%%%%%%%%%%%%%%%%%%%%%%%%

 One finds $e.g.$
 \begin{eqnarray}
\label{GG}
&&
f_0=0,~
f_1=1,~
f_2=\frac{r}{r_0},~
f_3= \frac{4}{9}+\frac{5r^2}{3r_0^2},~
f_4= \frac{55 r}{9 r_0}+\frac{35r^3}{3r_0^3},~
f_5=\frac{64}{225}+\frac{49 r^2}{15 r_0^2}+\frac{21r^4}{5r_0^4},~
\\
\nonumber
&&
g_0=1,~
g_1=\frac{r}{r_0},~
g_2=\frac{1}{3}+\frac{r^2}{r_0^2},~
g_3=\frac{r}{r_0}+\frac{5r^3}{3r_0^3},~
g_4=1+\frac{10 r^2}{3 r_0^2}+\frac{35 r^4}{3r_0^4},~
g_5=1+\frac{r}{r}+\frac{14 r^3}{3r_0^3}+\frac{21 r^5}{5r_0^5},~
\end{eqnarray}
The profile of the first four radial amplitudes 
are shown in Fig. (\ref{ex-sol})
(left panel).

However, the solution above 
is not fully satisfactory, since  
the derivatives of $R_\ell^{(+)}$ and $R_\ell^{(-)}$ do not match at the throat
(although $R_\ell^{(+)} (0)=R_\ell^{(-)} (0)$).
One finds
$e.g.$ 
$
\frac{dR_0^{(+)}}{dr}\big|_{r=0^+}
=-\frac{dR_0^{(-)}}{dr}\big|_{r=0^-}=\frac{1}{r_0},
$
$
\frac{dR_1^{(+)}}{dr}\big|_{r=0^+}
=-\frac{dR_1^{(-)}}{dr}\big|_{r=0^-}=-\frac{\pi}{2r_0},
$
and
$
\frac{dR_2^{(+)}}{dr}\big|_{r=0^+}
=-\frac{dR_2^{(-)}}{dr}\big|_{r=0^-}= \frac{4}{3 r_0}.
$
That is, the Klein-Gordon equation fails to be satisfied at the throat\footnote{
This can be seen by integrating the equation for the radial amplitude (\ref{eqR}) 
(multiplied with ($r^2+r_0^2$))
between $-\eps$ and $\eps$; 
then the $l.h.s.$ is just
 the difference between radial derivatives of $R_{\ell}^{\pm}$
evaluated at $\pm \eps$,
while the $r.h.s.$ vanishes as $\eps \to 0$ (since $R_\ell$ is continuous).
However, for a real scalar field ($i.e.$ $\omega=m=0$),
one can get a physically more reasonable picture
by supplementing (\ref{action}) with a 
boundary term 
\begin{equation}
\label{actionB}
S_B= - 2\sigma \int  d^4x  \sqrt{-g}\Phi \delta(r)~, 
\end{equation}
(where
$\sigma=  \frac{dR_\ell^{(+)}}{dr}\big|_{r=0^+}-\frac{dR_\ell^{(-)}}{dr}\big|_{r=0^-}$),
which acts as a thin shell source term for the Klein-Gordon equation located at the throat.
A similar situation occurs for WHs in Einstein-Gauss-Bonnet-dilaton theory, see Ref. \cite{Kanti:2011yv}.
}.

The only exception here is the $\ell=0$
mode, where the generic solution in
(\ref{ell}) (with the same choice for $c_1$, $c_2$ for all range of $r$)
has smooth derivatives at the throat, and
 approaches  a constant nonzero value 
at least in one of the asymptotic regions,
with 
$
R_0\to c_1+c_2\frac{\pi}{2}+O(1/r)
$
as $r\to \infty$, 
and
$
R_0\to c_1-c_2\frac{\pi}{2}+O(1/r) 
$
as $r\to -\infty$.
One remarks that 
 this radial function cannot vanish in both asymptotic regions.
The case of a $\ell=0$ solution 
with $R_0\to 0$   
as $r\to \pm  \infty$ 
is contained in
the general expression
(\ref{ell-gen}) 
(see also  Figure \ref{ex-sol}).
However, then the first derivative of $R_{0}$ is discontinuous at $r=0$.

%%%%%%%%%%%%%%%%%%%%%%%%%%%%%%%%%%%%%%%%%%%%%%%%%%%%%%%%%%%%%%%%%%
\subsection{The general case}
%%%%%%%%%%%%%%%%%%%%%%%%%%%%%%%%%%%%%%%%%%%%%%%%%%%%%%%%%%%%%%%%%%

Let us start with  
configurations satisfying the bound state condition
$\omega^2<\mu^2$, in which case  on may expect the existence
of smooth scalar clouds.
However, we have found that all properties of the $\ell>0$
solutions  
 discussed above hold also on this case.
The emerging picture can be summarized as follows.
When fixing   the constants $c_1$ and $c_2$,
both
functions  $H_1$ and $H_2$ in the general solution (\ref{gen-sol}) 
 diverge  for $|r|\to \infty$ and generic values of the parameters. 
However, it is possible to fine tune  $c_1$,  $c_2$
to get the solutions going to zero as $r \to \infty$,
but then they are divergent as $r\to -\infty$ (or viceversa).
In this case, the solutions are smooth at the throat, in particular with
$R_{\ell}(0^+)=R_{\ell}(0^-)$.

As with the  solutions in Section 3.1,
it is possible to construct a solution $R_{\ell}$ 
which goes to zero as $r\to \pm \infty$, by choosing a  different 
relation between $c_1$ and $c_2$ for each sign of $r$. 
However, then a discontinuous derivative of $R_{\ell}$ is unavoidable at $r=0$.

The absence of smooth  $C^1$ solutions with  $\omega^2<\mu^2$
can be seen from 
the following simple argument, which does not require an explicit form of the solution. 
After multiplying the eq. (\ref{eqR}) with $R_\ell$,
rearranging 
and integrating it, one finds that
 the solutions should satisfy the following identity:
\begin{equation}
\label{theorem}
  ( r^2 + r_0^2) R_\ell \frac{dR_\ell}{dr} \bigg|_{-\infty}^{\infty} =
\int_{-\infty}^{\infty} dr 
\left[	
( r^2 + r_0^2) \left( \frac{dR_\ell}{dr} \right)^2+
\left ( ( r^2 + r_0^2) (\mu^2 - \omega^2) +  {\ell (\ell+1)} \right) R_\ell^2
\right]	.
\end{equation}
The $l.h.s.$ vanishes identically (since  $R_\ell \sim e^{-\sqrt{\mu^2-\omega^2}|r|}/|r|$ as $r\to |\infty|$);
however, for $ \omega^2<\mu^2$ the $r.h.s.$ is strictly positive.
Thus we conclude the absence of physically reasonable scalar clouds in a Ellis-Bronnikov background
(note that for $\omega\neq 0$ this holds also for the $\ell=0$ mode).

\medskip 

The picture is rather different for 
 $\omega^2>\mu^2$,
in which case
the  solution is a smooth wave-like function. 
The asymptotic analysis reveals that the radial amplitude tends to zero as $|r\to \infty|$,
with the following far field regime behaviour\ 
$R_\ell \sim \frac{1}{|r|} (a_1\cos{\sqrt{\omega^2-\mu^2}r}+a_2\sin{\sqrt{\omega^2-\mu^2}r})$.  
At the throat, the solution possesses a power series expansion, the first term being
\begin{eqnarray}
\label{nte}
R_\ell(r)= c_1\left(
1+\left(\frac{1}{2}(\mu^2-\omega^2)+\frac{\ell(\ell+1)}{2r_0^2} \right)r^2
              +\dots \right)
	+c_2 \left (r+	\left(\frac{1}{4}(\mu^2-\omega^2)+\frac{(\ell-1)(\ell+2)}{4r_0^2} \right)r^3
             +\dots  \right)~~.~{~~}				
\end{eqnarray}
The profile of typical wave-like radial amplitudes is shown in Figure \ref{ex-sol} (right panel).
 In particular, one can notice that $H_1$ and $H_2$ are even and odd functions of $r$, respectively.
The inset there shows the general solution (the sum of $H_1$ and $H_2$), which possesses no parity.

%%%%%%%%%%%%%%%%%%%%%%%%%%%%%%%%%%%%%%%%%%%%%%%%%%%%%%%%%%%%%%%%%%
\section{A self-interacting scalar field:
 Q-balls in Ellis-Bronnikov }
%%%%%%%%%%%%%%%%%%%%%%%%%%%%%%%%%%%%%%%%%%%%%%%%%%%%%%%%%%%%%%%%%%
One may ask if the non-smooth behavior found for clouds
 in  the previous Section  can be cured by turning on the scalar field self-interaction.
The answer is positive, as shown by the existence of the following exact solution
  for a static, spherically symmetric real scalar field
with a  sextic potential\footnote{The solution (\ref{ex1})
has a  generalization with 
\begin{eqnarray}
\label{ex2}
\Phi=\frac{c_0}{(r^2+r_0^2)^{\frac{1}{k}} },~~~
 U(\Phi)= \lambda_1 \Phi^{k+2}+\lambda_2 \Phi^{2k+2},~~
{\rm with}~~ \lambda_1=\frac{4(2-k)}{k^2(k+2)c_0^k},~~\lambda_2=-\frac{4r_0^2}{k^2(k+1)c_0^{2k}}.
\end{eqnarray}
However, its mass is finite for  $k=1,2,3$ only.
 }
\begin{eqnarray}
\label{ex1}
\Phi=\frac{c_0}{\sqrt{r^2+r_0^2}},~~~
 U(\Phi)= \beta \Phi^6,~~{\rm where}~~\beta=-\frac{r_0^2 }{3c_0^4}.
\end{eqnarray}
This describes a smooth, even-parity configuration (with $\Phi'(0)=0)$,
possessing a finite mass
$E_\pm =\frac{2c_0^2\pi^2}{3r_0}$.
 However, this solution  does not seem
to possess generalizations with nonzero $\omega$ and $\mu$.

\medskip

Therefore, 
to better understand the issue of non-linear clouds 
in 
a WH background, we shall consider for the rest of this Section 
a more complicated scalar potential with
quadratic, quartic and sextic terms 
\begin{eqnarray}
\label{U} 
%U(|\Phi|) =  \l |\Phi|^2 \left( |\Phi|^4 -a |\Phi|^2 +b \right).
U(|\Phi|) =  \mu^2 |\Phi|^2-\lambda |\Phi|^4 +\beta |\Phi|^6.
\end{eqnarray}
with  $\lambda,\beta$ positive constants.
As discussed for the first time by Coleman in Ref. \cite{Coleman:1985ki},
this potential allows for nontopological soliton solutions
 in a flat spacetime background --{\it the Q-balls}. 
Such configurations have a rich structure and
found a variety
of physically interesting applications, see $e.g.$ the review work 
\cite{Radu:2008pp},
 \cite{Shnir:2018yzp}.
Here we shall focus on the simplest  Q-balls,
corresponding to spherically symmetric 
and spinning (even parity) configurations,
and show that the known flat space solutions
possess 
generalizations in a Ellis-Bronnikov WH 
background.
Moreover, as expected, the nonlinearities cure the pathological
behaviour of the scalar field at the throat, leading to smooth, finite mass solutions.

Also,
following the usual conventions in  the literature,
the numerical solutions reported here have been found for the following
parameters in the  potential (\ref{U}))
 \begin{eqnarray}
\mu^2=1, ~~\lambda=2,~~\beta=1.
 \end{eqnarray}

%%%%%%%%%%%%%%%%%%%%%%%%%%%%%%%%%%%%%%%%%%%%%%%%%%%%%%%%%%%%%%%%
\begin{figure}[ht!]
%\lbfig{rhfar}
\begin{center}
\includegraphics[height=.34\textwidth, angle =0 ]{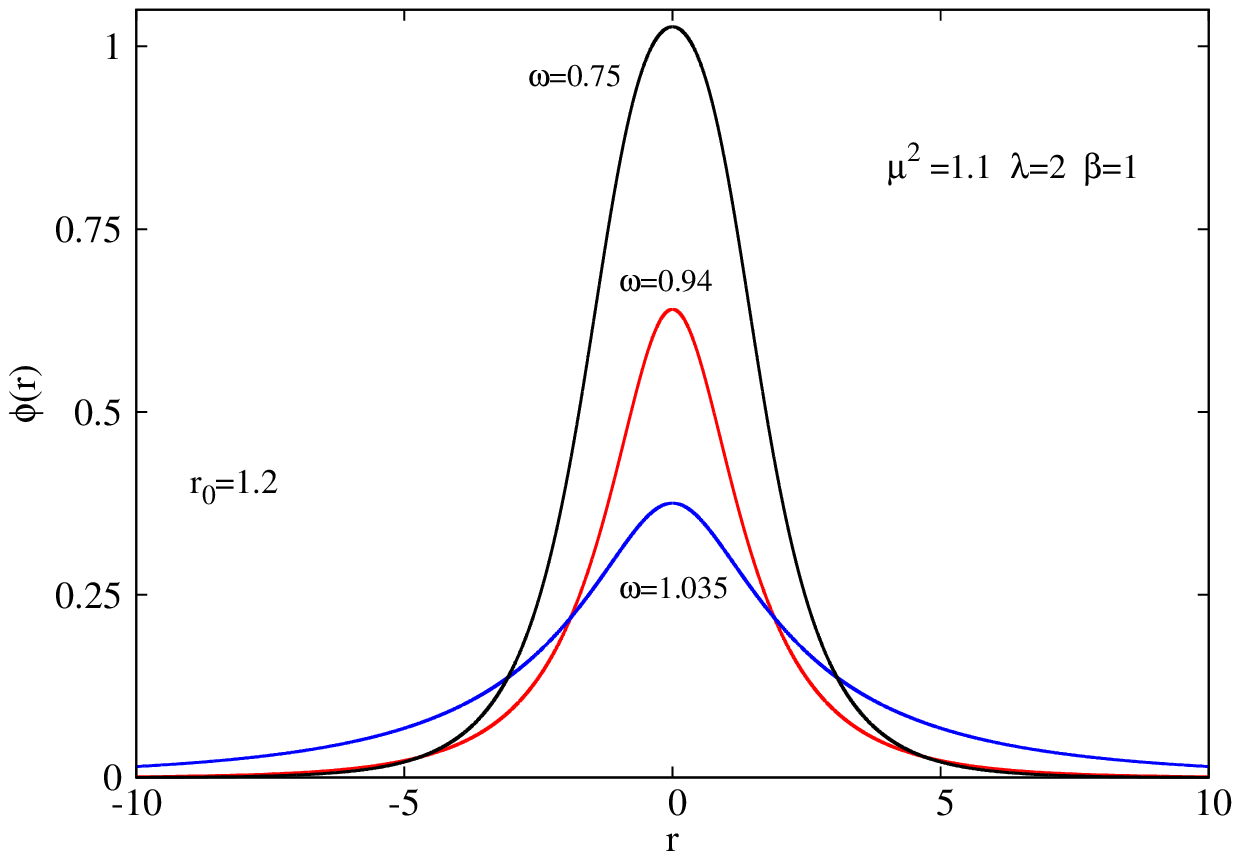}
\includegraphics[height=.34\textwidth, angle =0 ]{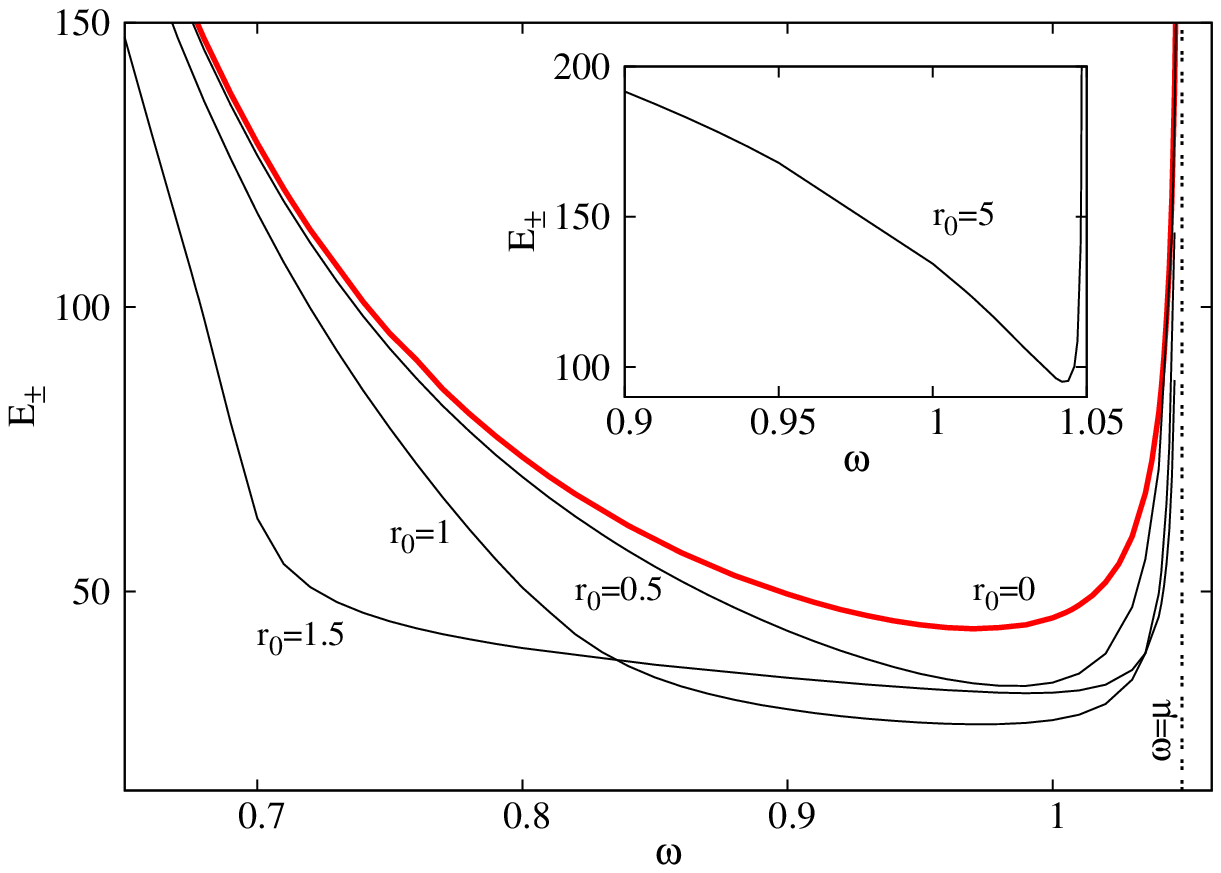}
\end{center}
\caption{
 {\it Left panel}:
 The profile of typical spherical Q-balls. 
 {\it Right panel}:
The mass-frequency diagram is shown for spherical Q-balls with several values of the throat radius $r_0$. 
}
\label{spherical}
\end{figure}
%%%%%%%%%%%%%%%%%%%%%%%%%%%%%%%%%%%%%%%%%%%%%%%%%%%%%%%%%%%%%%%%

%%%%%%%%%%%%%%%%%%%%%%%%%%%%%%%%%%%%%%%%%%%%%%%%%%%%%%%%%%%%%%%%%%
\subsection{Spherically symmetric Q-balls}
%%%%%%%%%%%%%%%%%%%%%%%%%%%%%%%%%%%%%%%%%%%%%%%%%%%%%%%%%%%%%%%%%%

The simplest Q-balls are spherically symmetric, with a scalar field Ansatz
\begin{eqnarray} 
\Phi=\phi(r) e^{-i \omega t}~,
\end{eqnarray}
where $\phi(r)$ is a real field amplitude 
(which is an even function of $r$, such that $E_+=E_-$),
and  $\omega>0$ is the frequency.
Close to the throat, one finds the following approximate solution,
\begin{eqnarray} 
 \phi(r)= \phi_0 + \phi_2 r^2+\dots,~~{\rm  where}~~
\phi_2 =\frac{1}{2}\phi_0 
(3 \phi_0^4-2 \phi_0^2 \lambda+\mu^2-\omega^2) ,
\end{eqnarray}
(note that no discontinuities occur at $r=0$),
while the field vanishes asymptotically,
$\phi\sim c_0 e^{-\sqrt{\mu^2-\omega^2}|r|}/|r|$.

The solutions connecting the above asymptotics 
are found numerically, 
the equation for  $\phi(r)$ being solved by
using a standard Runge-Kutta solver and implementing a shooting method.
Some results of the numerical integration are shown in Figure (\ref{spherical}).
In the left panel we display the profile of the radial amplitude for 
several frequencies (keeping fixed other parameters of the problem). 
The frequency-mass diagram 
is shown in the right panel,
for several values of the throat parameter $r_0$. 
One can see that the picture found for a Minkowski spacetime 
background is generic, the solutions 
existing for finite range of frequencies, $\omega_{min}<\omega<\mu$.
At the ends of this interval, the mass $E$ increases without bounds.
A similar behavior is found for the Noether charge\footnote{A  discussion  
(from a different perspective) 
of the spherically symmetric  Q-balls
in the  
 Ellis-Bronnikov WH background
can be found in 
Ref. \cite{Dzhunushaliev:2014bya}.} $Q$.

It is the non-linear self-interaction which circumvents the non-existence result 
(\ref{theorem}) (with $\ell=0$ and $R_0 \equiv \phi$).
Although the $l.h.s.$ there still vanishes for Q-balls,
 the term $(\mu^2-\omega^2)\phi^2$ on the $r.h.s.$
of that equation
  is replaced with 
 $(\mu^2-\omega^2) \phi^2-2\lambda \phi^4+3\beta \phi^6$,
which has no definite sign
(and in fact takes negative values for some range of $r$).
Also,
following the standard scaling arguments in the literature
\cite{Derrick:1964ww},
\cite{Herdeiro:2021teo},
 one can show that these solutions 
satisfy the virial identity
\begin{eqnarray} 
\label{virial}   
 \int_{-\infty}^\infty dr
\left[
(r^2-2r r_0-r_0^2) \phi'^2
+(3r^2-2r r_0+r_0^2)(U(\phi)-\omega^2 )
\right]=0.
\end{eqnarray}
One can see that, different from the flat spacetime case, the $r$-factor in front of the kinetic
term becomes negative for small enough $r$.  
This relation has been used as a further test of numerical accuracy.

%%%%%%%%%%%%%%%%%%%%%%%%%%%%%%%%%%%%%%%%%%%%%%%%%%%%%%%%%%%%%%%%
\begin{figure}[ht!]
%\lbfig{rhfar}
\begin{center}
\includegraphics[height=.34\textwidth, angle =0 ]{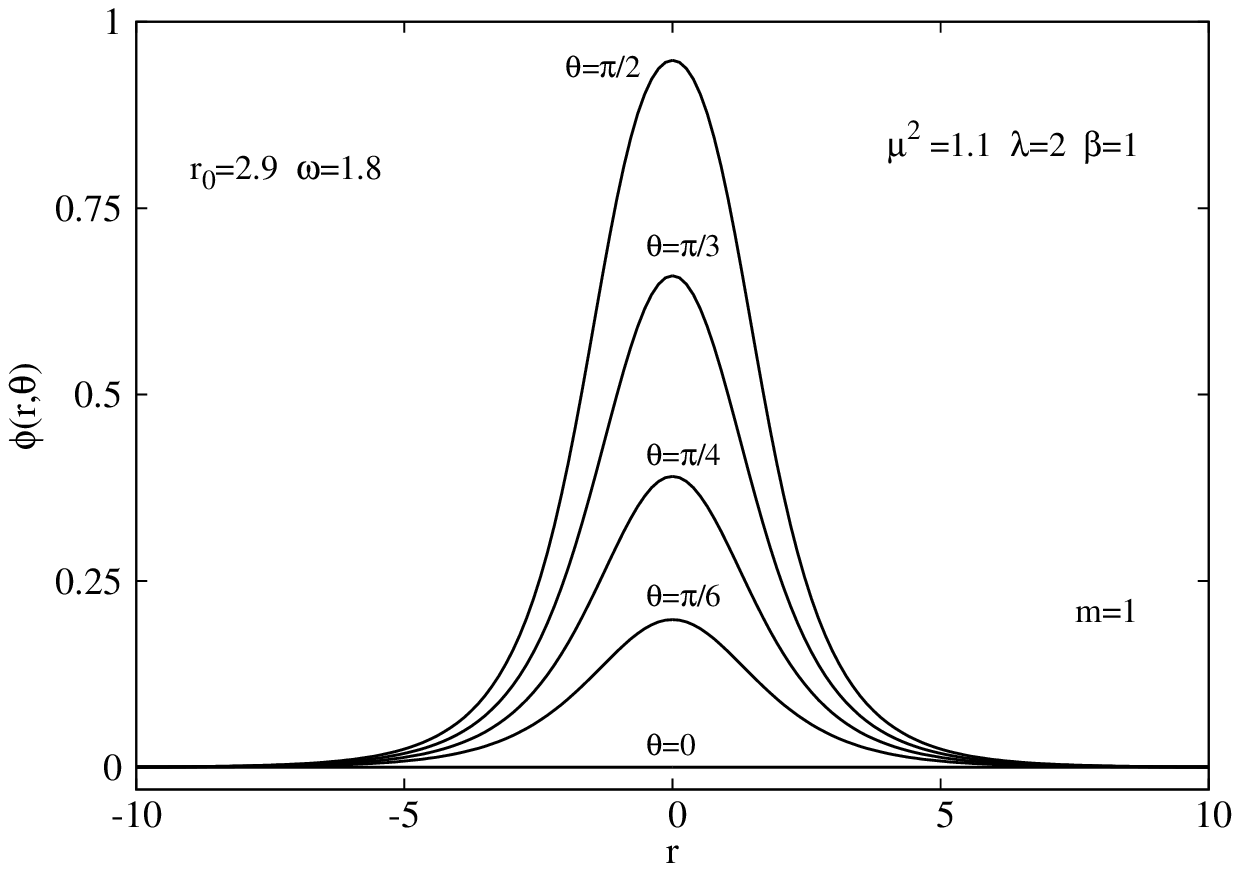}
\includegraphics[height=.34\textwidth, angle =0 ]{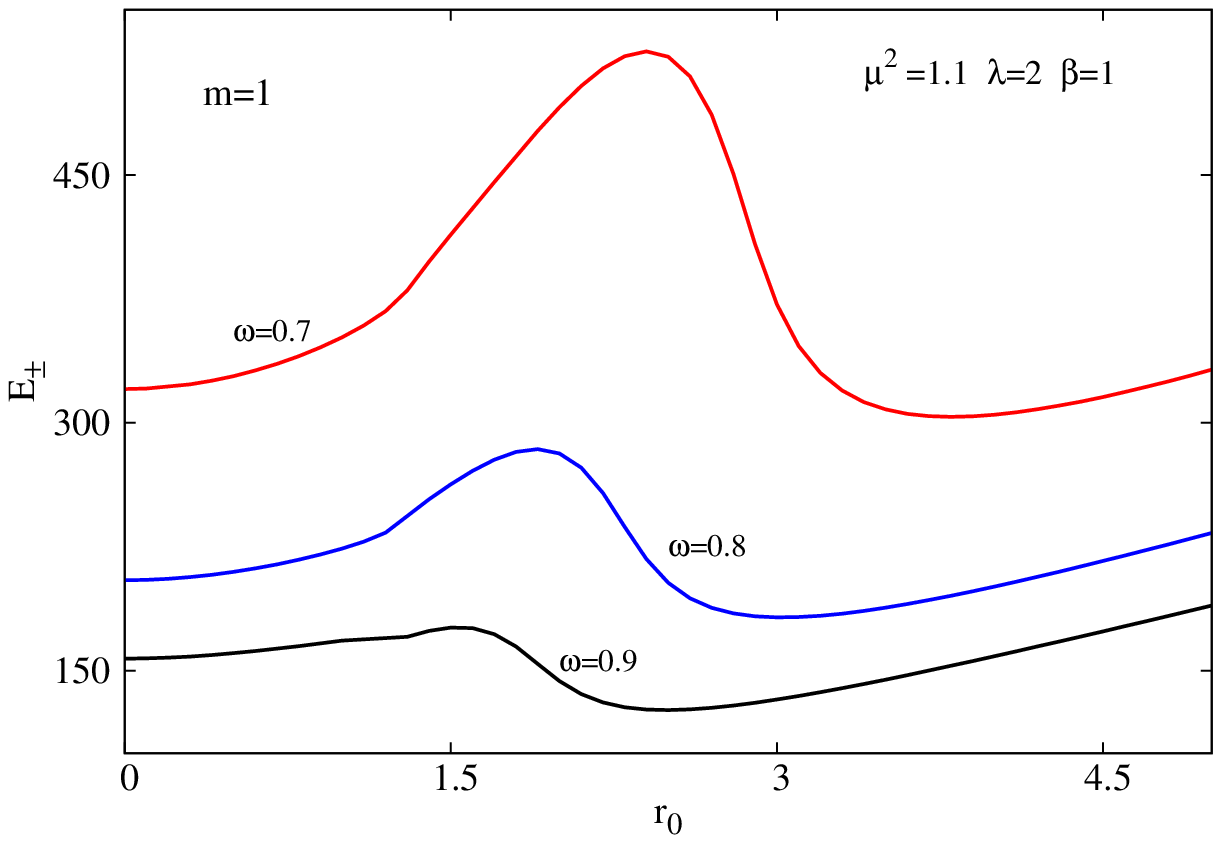}
\end{center}
\caption{
 {\it Left panel}:
 The scalar amplitude of a typical  spinning Q-ball is shown for several angular values. 
 {\it Right panel}:
The mass-frequency diagram is shown for spinning
 Q-balls with several values of the throat radius $r_0$.  
}
\label{spinning}
\end{figure}
%%%%%%%%%%%%%%%%%%%%%%%%%%%%%%%%%%%%%%%%%%%%%%%%%%%%%%%%%%%%%%%%

%%%%%%%%%%%%%%%%%%%%%%%%%%%%%%%%%%%%%%%%%%%%%%%%%%%%%%%%%%%%%%%%%%
\subsection{Spinning Q-balls}
%%%%%%%%%%%%%%%%%%%%%%%%%%%%%%%%%%%%%%%%%%%%%%%%%%%%%%%%%%%%%%%%%%
Solutions with a non-zero angular momentum exist as well, they 
being found for
a scalar field ansatz
\begin{eqnarray}
\label{scalar-ansatz}
\Phi=\phi(r,\theta)e^{i(m\varphi-\omega t)}~,
\end{eqnarray}
 where $\phi$ is a real function and $m=\pm 1,\pm 2 \dots$
is the azimuthal harmonic index. 
Note that  the $(t, \varphi)$-dependence of $\Phi$  occurs as a phase factor only,
such that the energy-momentum tensor depends on $(r, \theta)$, only 
(however, $m$ and $\omega$   still
enter  the expression of $T_{\mu\nu}$).
 These solutions possess a nonzero angular momentum density
$T_\varphi^t=2 m \omega \phi^2 =m j^t$,
and a total angular momentum
\begin{eqnarray}
\label{J}
 J_{\pm} =  \int_{\Sigma_\pm } d^3 x \sqrt{-g}~ T_\varphi^t .
 \end{eqnarray}
Interestingly,  the proportionality between angular momentum and Noether charge
found for a Minkowski spacetime background 
\cite{Volkov:2002aj},
\cite{Kleihaus:2005me}
still holds 
for a WH background,
 \begin{eqnarray}
\label{rel1}
J_\pm = m Q_\pm ,
\end{eqnarray}
such that  angular momentum  is still quantized.

With the ansatz ((\ref{metric}), (\ref{scalar-ansatz}), 
the Klein-Gordon equation (\ref{KG})
reduces to
\begin{eqnarray}
\label{KGn}
 \phi_{,rr}+\frac{1}{r^2+r_0^2}
\left(
\phi_{,\theta \theta}+2r \phi_{,r}+\cot \theta \phi_{,\theta}-\frac{m^2}{\sin^2 \theta }\phi
\right)
- (\mu^2 -\omega^2-2 \lambda \phi^2+3\beta \phi^4)\phi=0
\end{eqnarray}

We are interested in 
localized, particle-like solutions of this equation,
with a finite scalar amplitude $\phi$ and a regular energy density distribution.
In our approach, $Q$-clouds are found by solving the  
equation (\ref{KG1}) with suitable boundary conditions,
by using a professional package, 
based on the iterative Newton-Raphson method \cite{schoen},
the input parameters being $\{\omega,m; \mu,\lambda,\beta \}$.
The mass-energy and angular momentum are computed
from the numerical output.
The boundary conditions  result from the study of the solutions
on the boundary of the integration domain. 
The scalar field vanishes as $|r|\to \infty$, 
while the existence of a bound state requires $\omega <\mu$.
Also, the regularity of solutions 
 impose that the scalar field vanishes on the symmetry axis ($\theta=0,\pi$). 
This behavior holds also in the flat spacetime limit;
however, the boundary conditions at $r=0$ are different.
While in Minkowski spacetime the scalar field vanishes at the origin
 (as imposed by
the finiteness of various physical quantities), 
the condition for a WH  is 
$\partial_r \phi|_{r=0}=0$, while the field does not vanish at the throat.
As such, 
the configurations possess a reflection symmetry $w.r.t.$ to the throat, 
$\phi(-r)=\phi(r)$ and  
$E_+=E_-$,  
$J_+=J_-$.

 We restrict our study to 
 configurations which are invariant under a reflection in 
the equatorial plane
$\theta=\pi/2$.
Also, we shall restrict our study 
to configuration whose scalar amplitude $\phi(r,\theta)$ has no nodes.

The profile of a typical solution is displayed in Figure \ref{spinning} 
(left panel),
where the field amplitude $\phi$
is shown\footnote{Note that, due to the self-interaction, the field amplitude
can be thought as a superposition of infinite set of fundamental modes, 
%\begin{eqnarray}
$
 \phi(r,\theta)=\sum_{j=0}^\infty R_j (r)P_{2j+m}^m(\cos\theta)~,
$
%\end{eqnarray}
with  $P_j^m$ the associated Legendre functions.
} as a function of $r$
for several values of the polar angle $\theta$.
The $\omega$-dependence of solutions' mass 
is  qualitatively similar with that found in 
the spherically symmetric case, and we shall not exhibit it here.
We plot instead the mass dependence on the throat parameter $r_0$
for several values of the field frequency.
One can see that, rather counter-intuitive, this dependence is non-monotonic,
with the existence of local extrema.

%%%%%%%%%%%%%%%%%%%%%%%%%%%%%%%%%%%%%%%%%%%%%%%%%%%
\section{Further remarks. Conclusions}
%%%%%%%%%%%%%%%%%%%%%%%%%%%%%%%%%%%%%%%%%%%%%%%%%%%

For a wormhole (WH) geometry,
a two sphere possesses a minimal nonzero size,
which connects two asymptotically flat regions.
This property suggest that the usual $r=0$ divergence  of 
the solutions of a (linear) field theory model are absent in this case.
The main purpose of this paper was to investigate this aspect for 
the simplest case of a scalar field and a Ellis-Bronnikov wormhole  geometry.
 
Our results can be summarized as follows. 
For a free complex massive scalar field, the Klein-Gordon equation has a general exact solution that can be expressed in terms of Heun's functions, with two
distinct   classes of configurations.
For $\omega^2>\mu^2$ (with $\omega$ and $\mu$ the field's frequency and mass, respectively),
one finds wave-like solutions, which
propagates from one asymptotic region to another, being
 smooth everywhere.
The solutions with $\omega^2 \leq \mu^2$
are 'scalar clouds', the field amplitude vanishing asymptotically.
However, the Klein-Gordon equation fails to be satisfied at the throat, 
with a discontinuity in the radial derivative of the scalar field.
The only exception is found for the spherically symmetric mode with $\omega^2=\mu^2$,
which, in fact, has the same functional dependence on the radial coordinate as the phantom field that sources the  Ellis-Bronnikov solution \cite{Ellis}, \cite{Bronnikov}.
The pathological behaviour of the $\ell>0$ scalar clouds
strongly suggest the absence of multipolar deformation 
of the  Ellis-Bronnikov  WH, and can be viewed as a 'no-hair' theorem. 

In the second part of this work we addressed the question on how the field's
non-linearities may cure the scalar clouds' pathology found in the linear model.
Considering  a specific self-interacting potential which in flat spacetime
allows for particle-like solutions (the Q-balls), 
we have provided numerical evidence for the existence 
of smooth solitonic configurations in a Ellis-Bronnikov WH background.
Two different classes of solutions have been considered, corresponding to 
spherically symmetric and axially symmetric spinning  configurations 
which are invariant $w.r.t.$ a reflection in the equatorial plane.
However,  negative parity
solutions should also exist,
 their flat space limit
 being considered in 
\cite{Volkov:2002aj}, 
\cite{Kleihaus:2007vk}.
Also, 
on general grounds we predict the existence of a general tower of solutions
(Q-ball 'chains' and 'molecules' \cite{Herdeiro:2020kvf})
corresponding to regularized versions of 
the generic $(\ell,m,w)$-scalar clouds discussed in Section 3.1.

Another interesting question concerns the generality of the results
reported in this work.
Although a systematic work is clearly necessary, we expect some of the 
qualitative
results reported above to hold for a generic spherically symmetric WH, 
in particular those found for
linear waves and
 $Q$-balls.

%%%%%%%%%%%%%%%%%%%%%%%%%%%  
\section*{Acknowledgements}
%%%%%%%%%%%%%%%%%%%%%%%%%%%
The work of E. R. is supported by the Fundacao para a Ci\^encia e a Tecnologia (FCT) 
project UID/MAT/04106/2019 (CIDMA) and by national funds (OE), through FCT, I.P., in the scope of the framework contract foreseen in the numbers 4, 5 and 6
of the article 23, of the Decree-Law 57/2016, of August 29,
changed by Law 57/2017, of July 19. We acknowledge support  from the project PTDC/FIS-OUT/28407/2017 and PTDC/FIS-AST/3041/2020.  
 This work has further been supported by  the  European  Union's  Horizon  2020  research  and  innovation  (RISE) programmes H2020-MSCA-RISE-2015
Grant No.~StronGrHEP-690904 and H2020-MSCA-RISE-2017 Grant No.~FunFiCO-777740. 
E.R. would like to acknowledge networking support by the
COST Actions CA15117 {\sl CANTATA} 
and CA16104 {\sl GWverse}. JLBS gratefully acknowledges support by the
DFG Research Training Group 1620 {\sl Models of Gravity} and the DFG project BL 1553.
E.R. would like to acknowledge the hospitality of Ulm University (Raum 3102)
where a large part of this work has been done.

%%%%%%%%%%%%%%%%%%%%%%%%%%%%%%%%%%%%%%%%%%%%%%%%%%%%%%%%%%%%%%%%%%%%%%%%%%%%%%  
\begin{small}

 \end{small}
%%%%%%%%%%%%%%%%%%%%%%%%%%%%%%%%%%%%%%%%%%%%%%%%%%%%%%%%%%%%

\end{document}